\documentclass{article}

\usepackage{PRIMEarxiv}

\usepackage[utf8]{inputenc} 
\usepackage[T1]{fontenc}    
\usepackage{hyperref}       
\usepackage{url}            
\usepackage{booktabs}       
\usepackage{amsfonts}       
\usepackage{nicefrac}       
\usepackage{microtype}      
\usepackage{lipsum}
\usepackage{fancyhdr}       
\usepackage{graphicx}       
\graphicspath{{media/}}     

\usepackage{amsmath,amssymb,amsfonts}
\usepackage{algorithmic}
\usepackage{graphicx}
\usepackage{textcomp}
\usepackage{xcolor}
\def\BibTeX{{\rm B\kern-.05em{\sc i\kern-.025em b}\kern-.08em
    T\kern-.1667em\lower.7ex\hbox{E}\kern-.125emX}}
    
\usepackage[utf8]{inputenc} 
\usepackage[T1]{fontenc}    
\usepackage{hyperref}       
\usepackage{url}            
\usepackage{booktabs}       
\usepackage{amsfonts}       
\usepackage{nicefrac}       
\usepackage{microtype}      
\usepackage{xcolor}         
\usepackage{multirow}       
\usepackage{graphicx}       
\usepackage{subcaption}
\usepackage{chngcntr}
\usepackage{amsmath}
\usepackage{amssymb}
\usepackage{biblatex}
\usepackage[english]{babel}
\usepackage{blindtext}
\usepackage{footmisc}
\usepackage{hyperref}
\usepackage[justification=centering]{caption} 
\usepackage{threeparttable} 
\def\BibTeX{{\rm B\kern-.05em{\sc i\kern-.025em b}\kern-.08em
    T\kern-.1667em\lower.7ex\hbox{E}\kern-.125emX}}

\title{TradingGPT: Multi-Agent System with Layered Memory and Distinct Characters for Enhanced Financial Trading Performance
}

\author{
  Yang Li, Yangyang Yu, Haohang Li, Zhi Chen, Khaldoun Khashanah \\
    School of Business, Stevens Institute of Technology \\
    Hoboken, NJ, United States \\
  \texttt{\{yli269, yyu44, hli113, zchen100, kkhashan\}@stevens.edu} \\
}

\begin{document}
\maketitle

\begin{abstract}
Large Language Models (LLMs), prominently highlighted by the recent evolution in the Generative Pre-trained Transformers (GPT) series, have displayed significant prowess across various domains, such as aiding in healthcare diagnostics and curating analytical business reports. The efficacy of GPTs lies in their ability to decode human instructions, achieved through comprehensively processing historical inputs as an entirety within their memory system. Yet, the memory processing of GPTs does not precisely emulate the hierarchical nature of human memory, which is categorized into long, medium, and short-term layers. This can result in LLMs struggling to prioritize immediate and critical tasks efficiently. To bridge this gap, we introduce an innovative LLM multi-agent framework endowed with layered memories. We assert that this framework is well-suited for stock and fund trading, where the extraction of highly relevant insights from hierarchical financial data is imperative to inform trading decisions. Within this framework, one agent organizes memory into three distinct layers, each governed by a custom decay mechanism, aligning more closely with human cognitive processes. Agents can also engage in inter-agent communication and debate. In financial trading contexts, LLMs serve as the decision core for trading agents, leveraging their layered memory system to integrate multi-source historical actions and market insights. This equips them to navigate financial changes, formulate strategies, and debate with peer agents about investment decisions. Another standout feature of our approach is to enable agents with individualized trading characters, which enrich the diversity of their highlighted essential memories and improve decision-making robustness. By leveraging agents' layered memory processing and consistent information interchange, the entire trading system demonstrates augmented adaptability to historical trades and real-time market cues. This synergistic approach guarantees premier automated trading with heightened execution accuracy.
\end{abstract}


\section{Introduction}
As the influx of diverse data streams continues to rise, there is a growing need for individuals to effectively harness information. This trend is particularly pronounced in the realm of finance, where traders must consider multiple sources to inform their investment decisions. In light of this demand, researchers design intelligent trading robot-agents that can synthesize and interpret data objectively\cite{shin2019automatic, huang2022mspm}. These robot-agents harness diverse machine algorithms, assimilate a broader spectrum of data, autonomously refine trading strategies via methodical planning, and even potentially collaborate \cite{liu2020adaptive}. Here, we introduce an advanced LLM-powered multi-agent trading agent framework, supported by layered memories and customized characters. By employing a collaborative multi-agent system and capturing the intricate market dynamics from varied perspectives, this approach significantly enhances automated trading outcomes. This approach substantially elevates the performance of automated trading by fostering collaborative interactions among agents and capturing the intricate dynamics of the market from diverse perspectives.

Previous studies have introduced multi-agent trading algorithms that employ machine learning techniques, such as reinforcement learning and have reported significant performance outcomes \cite{huang2022mspm}. Yet, these methods exhibit limitations in precisely identifying, representing, and emulating crucial components of trading systems. This includes aspects like agents' memory archives and the evolving social interplay among agents.

LLMs, with a particular focus on their recent advancements, such as the Generative Pre-trained Transformer (GPT), have demonstrated remarkable effectiveness in enhancing human decision-making across various domains \cite{openai2023gpt4}. Notably, a growing body of research has focused on harnessing this technology to make informed trading decisions for stocks and funds by continuously interacting with financial environment information \cite{yang2023fingpt, wu2023bloomberggpt}. While current financial LLM applications predominantly operate within single-agent systems based on textual uni-modality, their immense potential to elevate trading performance is becoming increasingly evident. Moreover, these financial agent systems make trading decisions relying solely on pre-trained LLMs or a memory system processing received information streams as an entirety. This can lead to a challenge for LLMs in efficiently prioritizing immediate and critical memory events for optimized trading. 

Park et al. \cite{park2023generative} recently introduced a generative agent framework aiming to enhance the efficient retrieval of critical events from agents empowered by LLMs. This structure comprises several agents, each distinguished by separate memory streams and unique character profiles configured by LLMs. Each agent, owning its seed memories, not only tracks its actions but also monitors other agents and environmental behaviors. Faced with a task, agents sift through memory segments to input into the language model, ranking them by recency, significance, and relevance. By archiving an agent's experiences, the system integrates individual weighted memories and the nuances of group dynamics. As a result, agents can collaboratively strategize, leveraging their collective knowledge. Moreover, Du et al. \cite{du2023improving} presented a debate mechanism for LLM agents, emphasizing enhanced cooperative decision-making through debate phases in inter-agent memory interactions. These advancements align the LLM-driven multi-agent system more with human memory structures, paving the way for a more adept financial automated trading system.

Leveraging the capabilities of LLMs, we propose a novel trading agent framework, "TradingGPT". It offers a realistic scenario simulation through the integration of the trader's layered memory streams and character analysis. This framework is characterized by remarkable self-enhancement ability and performance to conduct automated trading and optimal execution. The primary contributions of our work include: 

\textbf{This represents a pioneering multi-agent trading system that integrates memory streams and debate mechanisms,} anchored on LLMs. Building on Park et al.'s weighted memory mechanisms, our system innovatively categorizes the agent's memories into short-term, middle-term, and long-term layers, which are closely aligned with the structure of the human cognitive system. We adapt this layered memory framework to the financial trading system, equipping agents to reflect on past and present events, derive insights from trading performance, and leverage collective wisdom for future decisions. This approach improves the system's robustness.

\textbf{This marks the debut of the LLM agent trading system that incorporates the character design.} The design assigns agents with different varying risk preferences, such as risk-seeking, risk-neutral, and risk-averse, and various investment subscopes across industries. This design enables these collaborative agents to resonate more with human intuition and possess the potential to uncover latent market opportunities.

\textbf{Our trading system also integrates real-time multi-modal data from diverse information sources}, offering a comprehensive view of the financial landscape by encompassing both macro and micro perspectives, as well as historical trading records. With updates available on both daily and minute-by-minute frequencies, our system ensures prompt reactions to daily trades and offers the capability for high-frequency trading.

In this paper, we commence with an in-depth exposition of TradingGPT. We then present multi-modal datasets for the effective training of TradingGPT. We methodically evaluate the pivotal components of the system, illustrating their ability to yield notable results. We prospect that, when deployed on representative fund firms like ARK, TradingGPT will markedly outperform other automated trading strategies.

\section{Related Work}

\subsection{Large language models (LLMs)}
The evolution of LLMs has reshaped artificial intelligence and natural language processing. From foundational embeddings like Word2Vec \cite{goldberg2014word2vec} and GloVe \cite{pennington2014glove}, the field advanced with the introduction of BERT \cite{devlin2018bert}. Today, the new-generation LLMs, like Generative Pre-trained Transformer series (GPTs) \cite{radford2018improving, openai2023gpt4} and Large Language Model Meta AI (Llamas) \cite{touvron2023llama}, demonstrate expressive proficiency across diverse applications.

\subsection{Generative agent system with memory streams and customized character design}
Park et al. \cite{park2023generative} introduced generative agents' memory streams and innovatively employed character design concepts from gaming, expanding LLM capabilities for the multi-agent system \cite{riedl2012interactive}. In their design, agents display human-like behaviors while retaining individual characters. They dynamically interact with peers and their environment, forging memories and relationships. Moreover, these agents coordinate collaborative tasks through natural language, creating a captivating fusion of artificial intelligence and interactive design.

\subsection{Multi-agent debate mechanism}
Du et al. \cite{du2023improving} introduced a debate mechanism leveraging multiple language models in a multi-agent system. Within this framework, various model instances propose debate and collaboratively converge to a unified answer. This approach bolsters mathematical and strategic reasoning while enhancing the factual accuracy of the generated content.

\begin{figure}[htbp]
\centering
\includegraphics[width=12.2cm]{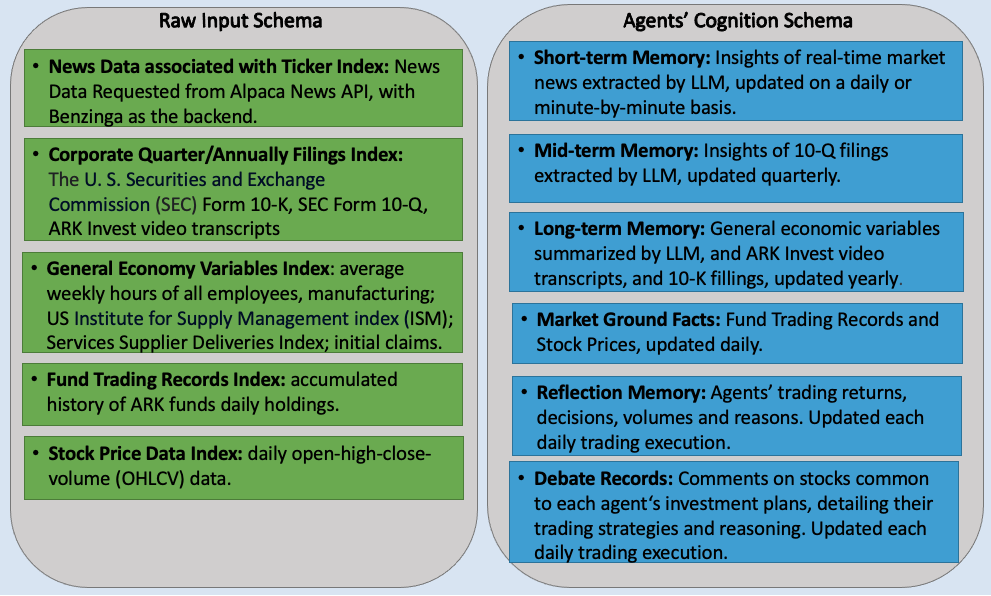}
\caption{TradingGPT Data Warehouse.} 
\label{fig2:db}
\end{figure}

\section{Dataset and Database Structure}
For TradingGPT's development,  we systematically integrated an extensive array of multi-modal financial data from August 15, 2020, to August 15, 2023. These datasets were sourced from financial databases and APIs, exemplified by the Databento Stock Price Database, Alpaca News API, publicly available daily holdings history records from ARK, etc. This data serves two purposes: (a) to formulate multi-layer memories for agents, and (b) to train, guide, and back-test the agents using ARK funds' historical trading records, refining their trading decisions and actions. In our study, we employed FAISS\cite{johnson2019billion}, an open-source vector database, due to its capacity to store data as high-dimensional vectors, enabling semantic searches based on exact matches. Two primary reasons informed our decision: (a) The majority of our data, including audio transcriptions from ARK Invest videos (translated to texts via the Whisper API), benefits from FAISS's unique underlying structure to fast query data. (b) FAISS's compatibility incorporating OpenAI and efficient computation of cosine similarities for specific tickers.
the Raw Input schema. This data is then channeled into the Agents' Cognition Schema, guided by both the system's foundational logic and LLM-agent processing. A comprehensive schema structure is in Figure.~\ref{fig2:db}.

\section{Proposed Method}
Our methodology integrates LLM across multiple facets of the trading agent workflow. Details and associated notation are provided in the subsequent sections.

\subsection{Trading Agents Layered Generative Memory Formulation}
\label{mem_form}
In our LLM-based trading system, agents autonomously manage their actions and memory trajectories, engaging in communication and deliberation as needed.

\subsubsection{Layered-memory structure}
\label{layered_mem}

Each agent within TradingGPT discerns and categorizes perceived information into three distinct memory layers: long-term, middle-term, and short-term. Compared to the approach of extracting key insights through the computation of ranked retrieval scores from all memories in the generative agent system \cite{park2023generative}, this layered memory approach introduces a more nuanced ranking mechanism for retrieving crucial events from individual layers. This closely aligns with the human cognition proposed by Atkinson et al.\cite{atkinson1968human}. Our framework initially categorizes memories into separate lists for each layer, guided by predefined rules tailored to specific situations and the nature of events. Subsequently, within each memory layer, we leverage three crucial metrics, inspired by the work of Park et al. - recency, relevancy, and importance - to establish the hierarchical arrangement of events within an agent's memory. However, we have reconstructed their mathematical representations to attain a more logical and advanced formulation. 

For a memory event $E$ within the memory layer $i \in \{\text{short, middle, long}\}$, upon the arrival of a prompt $P$ from the LLM, the agent computes the recency score $S_{\text{Recency}}^{E}$ as per Equation.\ref{eqn:eq1}. This score inversely correlates with the time difference between the prompt's arrival and the event's memory timestamp, aligning with Ebbinghaus's forgetting curve on memory decay \cite{murre2015replication}. $Q_i$ Equation.\ref{eqn:eq1} represents the stability term, employed to control the memory decay rates across layers. A higher stability value in the long-term memory layer compared to the short-term layer suggests that memories persist longer in the former. The relevancy score $S_{\text{relevancy}}^{E}$ represents the cosine similarity between the embedding vectors for the textual content of the memory event $\mathbf{m_{E}}$ and the prompt query $\mathbf{m_{P}}$. The importance score $S_{\text{Importance}}^{E}$ is determined using a uniform piecewise function as described in Equation.\ref{eqn:eq3}, adhering to the relationship $c_{\text{short}}< c_{\text{middle}}< c_{\text{long}}$. After normalizing their values to the [0,1] range using min-max scaling, these scores, $S_{\text{Recency}}^{E}$, $S_{\text{Relevancy}}^{E}$ and $S_{\text{Importance}}^{E}$ are linearly combined to produce the final ranking score $\gamma_i^{E}$ for each memory layer in the Equation.~\ref{eqn:eq4} (equivalent to retrieval score in the study of Park et al.). In our setup, the ranking score thresholds, $\gamma_i^{E}$, are 80 for long-term, 60 for middle-term, and 40 for short-term memory. Events scoring below 20 are removed.

\begin{equation}
  \begin{split}
& S_{\text{Recency}}^{E} = e^{-\frac{\delta^{E}}{Q_i}}  \quad  \delta^{E} = t_{\text{P}} - t_{E} \\
 \end{split}
 \label{eqn:eq1}
  \end{equation}
, where $Q_{\text{long}} = 365$ for long-term, $Q_{\text{middle}} = 90$ for middle-term, and $Q_{\text{short}} = 3$ for short-term events.

\begin{equation}
  \begin{split}
& S_{\text{Relevancy}}^{E} = \frac{\mathbf{m_{E}} \cdot \mathbf{m_{P}}}{\|\mathbf{m_{E}}\|_2 \times \|\mathbf{m_{P}}\|_2}
 \end{split}
 \label{eqn:eq2}
  \end{equation}

\begin{equation}
  \begin{split}
& S_{\text{Importance}}^{E} = \begin{cases} 
c_{\text{short}} & \text{if short-term memory}\\
c_{\text{middle}} & \text{if middle-term memory}  \\
c_{\text{long}} & \text{if long-term memory} 
\end{cases}
 \end{split}
 \label{eqn:eq3}
  \end{equation}
, where $c_{\text{short}}, c_{\text{middle}}$ and $c_{\text{long}}$ are all constants.

\begin{equation}
\gamma_i^{E} = \alpha_i^{E} \times S_{\text{Recency}_{i}}^{E} +\beta_i^{E} \times S_{\text{Relevancy}_{i}}^{E} + \lambda_i^{E} \times S_{\text{Importance}_{i}}^{E}
 \label{eqn:eq4}
  \end{equation}
where each memory event is only associated with one score, as it can only belong to one of the memory layers.

\begin{figure*}[htbp]
\centering
\includegraphics[width=16.8cm]{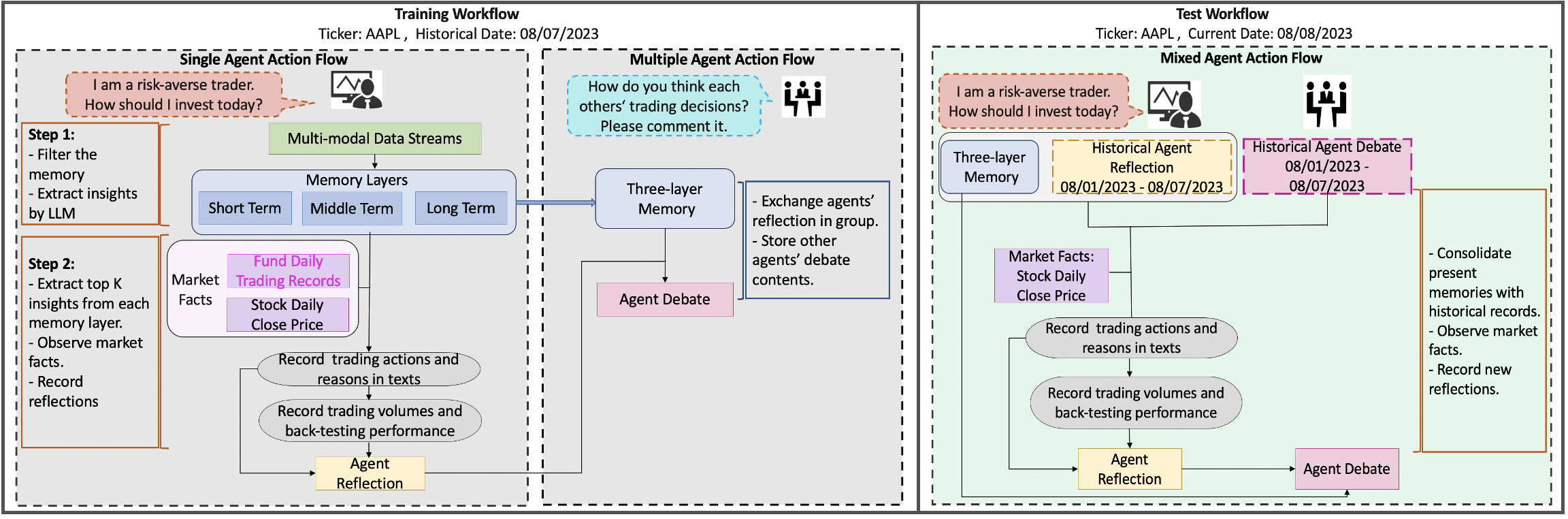}
\caption{TradingGPT training and test workflow. }\footnotemark
\label{fig1:training_logic}
\end{figure*}

To ensure dynamic interactions across memory layers, we define upper and lower thresholds for memory event ranking scores in each layer. We also utilize an add-counter function to boost the scores of events that are triggered by trading executions resulting from significant trading profits and losses. This promotes frequent events to transition from short-term to potentially longer-term memory, enhancing their retention and recall by agents. The hyperparameters $\alpha_i^{E}$, $\beta_i^{E}$, and $\lambda_i^{E}$ exhibit variations across different layers. The transferable layered memory system allows the agents to capture and prioritize crucial memory events by considering both their types and frequencies when conducting queries.

\subsubsection{Memory formulated by individual experience}
\label{single_mem_form}

In the trading paradigm, macro-level market indicators are stored in the long-term memory, quarterly investment strategies are allocated to the mid-term memory, and daily investment messages are channeled into the short-term memory. These three memory classes constitute the initial structure within the Agents' Cognition Schema of our data warehouse in Figure.~\ref{fig2:db}. In our trading system, agents make informed trading decisions relying on the outcomes of two distinct workflows: the single-agent workflow and the multi-agent workflow, as depicted on the left side of Figure~\ref{fig1:training_logic}. 

In the single-agent workflow, when presented with a specific stock ticker, agents' LLM core generates evaluations and reflections, which encompass trading recommendations and the reasons behind them, based on the essential events retrieved from their layered memory. Subsequently, the agent can proceed to execute trading actions in accordance with these generated insights. The key features that empower our system are (a) Immediate reflection: Conducted daily, this mechanism allows agents to consolidate top-ranked events of each memory layer and market facts, such as daily stock prices and ARK fund trading records. Using LLM and specific prompts, agents generate five trading recommendations: “significantly increase position”, “slightly increase position", “hold”, “slightly decrease position”, and “significantly decrease position”, with its justification. Each option is associated with a predetermined trade value. which can be adjusted to suit the business scale represented by the agents. Additionally, this reflection captures the agent's trade volumes and returns. (b)Extended reflection: This provides a broader performance overview over a designated period, like a week. It includes stock prices, the agent's trading trends, and self-evaluation. The immediate reflection guides trade execution directly, while the extended reflection acts as a supplementary reference for recalling recent investment transactions. Both types of reflections are stored in the Agents' Cognition Schema's reflection index, as shown in Figure~\ref{fig2:db}, distinguished by a specific flag.

\subsubsection{Memory gained by interacting with other agents}
\label{debate}
For stocks that appear in multiple agents' trading portfolios, TradingGPT enables inter-agent dialogue via a debate mechanism. This mechanism encourages collaboration between agents typically specializing in distinct sectors, with the goal of optimizing trading outcomes. Within these debates, agents present their top-K layered memories as well as immediate reflections, encompassing recommendations, trade values, volumes, and returns, inviting feedback from their peers. All feedback is subsequently stored in the debate class of the Agents' Cognition Schema, tagged with the receiver's index, as shown in Figure.~\ref{fig2:db}.
\footnotetext[1]{Data entities without specific timestamps are extracted as per the date displayed at the top of the plots.}

\subsection{Design of Training and Testing Workflows}
The distinct design of our training and testing workflows is crucial for curating valuable past memory events and strategizing optimal future trading actions. 

\subsubsection{Training} 

The training process is twofold: a single-agent workflow followed by a multi-agent phase, as detailed in the left section of Fig.~\ref{fig1:training_logic}. In the single-agent phase, the LLM-driven agent is prompted with key data like stock ticker, date, and trader characters. Using this context, it evaluates top-K-ranked memories across each layer to derive preliminary investment signals, where K is a predefined hyperparameter. The LLM then synchronizes and analyzes these signals with market data, such as daily records from fund firms like ARK and stock closing prices, leading the agent to formulate an immediate reflection and trade accordingly. Subsequently, the agent collaborates in the multi-agent phase, joining debates with agents trading the same stock from varied sectors on that day (refer to \ref{debate}). 

\subsubsection{Test}
The testing process, illustrated in the right section of Figure.~\ref{fig1:training_logic}, blends single-agent and multi-agent operations. Both individually processed memories and insights from inter-agent exchanges are concurrently inputted into the LLM to inform trading decisions. Key differences from the training phase include:
(a) During testing, agents operate without the guidance of trading records from the representative fund firm, relying solely on daily stock prices as market facts.
(b) Time series patterns of prior training reflections and debates, covering a week in our setup, act as auxiliary references in the absence of substantial market ground truths, as noted in (a). Other aspects of the test workflow align with the training phase.

\begin{figure}[htbp]
\centering
\includegraphics[width=7.0cm]{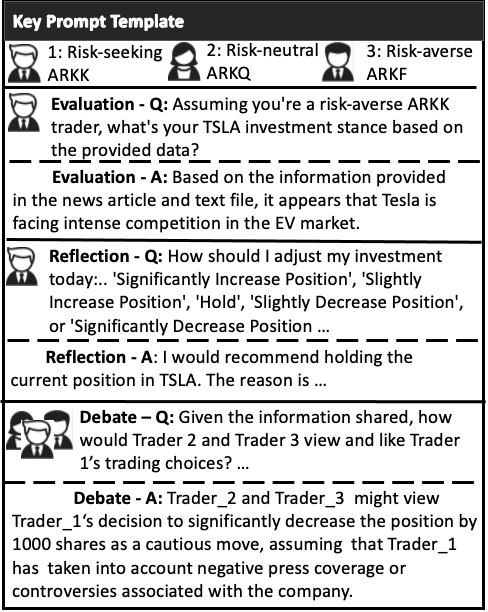}
\caption{Prompt template for key steps of TradingGPT workflow.}
\label{fig3:prompt}
\end{figure}

\section{Current Stage And Future work}

Our research consists of two phases: prompt design and ablation studies. We've crafted efficient LLM prompts using  GPT3.5 turbo as the backbone. Examples of prompts that encapsulate the necessary insights for each phase of the TradingGPT training and testing workflow. The specific design of these prompts is illustrated by examples in Figure.~\ref{fig3:prompt}.

With our established prompt template, we're poised to undertake ablation studies to assess the trading efficacy of agent systems based on various backbone models. This will involve comparisons within LLMs, such as GPT3.5 turbo versus CodeLlama 34B, and against models like multi-agent reinforcement learning. The training phase will utilize data spanning from August 15, 2020, to February 15, 2023, while the testing phase will extend until August 15, 2023. We'll assess performance using financial metrics like cumulative trade returns, volatility, and the Sharpe Ratio (see \ref{single_mem_form}).

Harnessing an innovative multi-layer memory system and character design, our main goal is to establish a state-of-the-art LLM-based multi-agent automated trading system adaptable to various LLMs as its core. This system aspires to achieve superior trading performance over other leading trading agent systems by emulating human traders' cognitive behaviors and ensuring responsiveness in the constantly changing market scenario. We also posit that this LLM-based multi-agent design can improve working efficiency and collaborative performance in artificial systems across diverse sectors. Potential applications range from character development in video games to the creation of robo-consultants in business, healthcare, and technology domains.

\printbibliography

\end{document}